\documentclass[12pt,prd,showpacs,nofootinbib,superscriptaddress]{revtex4}

\usepackage[T1]{fontenc}
\usepackage[english]{babel}

\usepackage[latin1]{inputenc}

\usepackage{amssymb}

\usepackage{graphicx}

\usepackage{amsmath}

\usepackage{mathrsfs}

\usepackage{bm}

\usepackage{dsfont}

\usepackage{longtable}

\usepackage{color}

\usepackage{anysize}

\def\be{\nopagebreak[3]\begin{equation}}
\def\ee{\end{equation}}
\def\ba{\nopagebreak[3]\begin{eqnarray}}
\def\ea{\end{eqnarray}}

\newcommand{\teta}{\rlap{\lower2ex\hbox{$\,\tilde{}$}}\eta{}}

\definecolor{rojo}{rgb}{1,0,0}
\definecolor{azul}{rgb}{0,0,1}

\begin{document}

\title{Black hole state degeneracy in Loop Quantum Gravity}

\author{Iv\'an Agull\'o}
\email{Ivan.Agullo@uv.es}
\affiliation{{\footnotesize Department of Physics, University of
Maryland, College Park, Maryland 20742}}
\affiliation{{\footnotesize Departamento de F\'\i sica Te\'orica and IFIC, Centro
Mixto Universidad de Valencia-CSIC. Facultad de F\'\i sica, Universidad de Valencia,
Burjassot-46100, Valencia, Spain.}}

\author{Jacobo D\'\i az-Polo}
\email{Jacobo.Diaz@uv.es}
\affiliation{{\footnotesize Departamento de Astronom\'\i a
y Astrof\'\i sica, Universidad de Valencia, Burjassot-46100,
Valencia, Spain}}
\affiliation{{\footnotesize Instituto de Matem\'aticas, Universidad Nacional Aut\'onoma de M\'exico,\\
A. Postal 61-3, Morelia, Michoac\'an 58090, M\'exico}}

\author{Enrique Fern\'andez-Borja}\email{Enrique.Fernandez@uv.es}
\affiliation{{\footnotesize Departamento de F\'\i sica Te\'orica and IFIC, Centro
Mixto Universidad de Valencia-CSIC. Facultad de F\'\i sica, Universidad de Valencia,
Burjassot-46100, Valencia, Spain.}}
\affiliation{{\footnotesize Departamento de Astronom\'\i a
y Astrof\'\i sica, Universidad de Valencia, Burjassot-46100,
Valencia, Spain}}

\begin{abstract}

\begin{center}
{\bf Abstract}
\end{center}

\noindent
The combinatorial problem of counting the black hole quantum states within the Isolated Horizon framework in Loop Quantum Gravity is analyzed. A qualitative understanding of the origin of the band structure shown by the degeneracy spectrum, which is responsible for the black hole entropy quantization, is reached. Even when motivated by simple considerations, this picture allows to obtain analytical expressions for the most relevant quantities associated to this effect.

\end{abstract}


\maketitle

\section{Introduction}
\label{intro}

In 1974 S.W.~Hawking \cite{hawk1} established that black holes behave
like black bodies in the thermodynamical sense. This remarkable work
provides a clear evidence that the similarity between the laws of black
hole mechanics \cite{BCH} and the ordinary laws of thermodynamics is
much more than a mere mathematical analogy. This physical analogy is
summarized in the Generalized Second Law \cite{Bekenstein74}, which
endows black holes with physical entropy in a pure thermodynamical
sense. Any quantum gravity theory proposal has to provide the
microscopic degrees of freedom that
account for that entropy.\\
Loop Quantum Gravity (LQG) \cite{ash-lew, perez, rovelli, thiemann}
offers a detailed description of the black hole horizon
quantum states. Black holes within LQG are treated in an
effective way in the Isolated Horizon framework introduced by
Ashtekar \emph{et al} \cite{ABK}. In this framework the horizon
quantum degrees of freedom are described by a $U(1)$ Chern-Simons
gauge theory and fluctuate independently from the ones of the bulk, giving rise to the black hole entropy.
At present, two inequivalent proposals \cite{DL,GM} for characterizing the black hole degrees of freedom have received most of the attention. It is interesting that, within both of them, the problem of computing
the black hole entropy can be reduced to a well defined
\emph{combinatorial} problem.

To exactly solve this combinatorial problem is, however, a rather non trivial task and,
in order to obtain analytical solutions, some approximations have to be
made. In particular the large area approximation
permits to perform an analytic counting of the black hole microstates
\cite{M,GM,GMdist}. Using this approximation the theory reproduces
the semiclassical proportionality relation between entropy and area and gives an additional logarithmic term with a $-1/2$ coefficient,
\be
\label{SA}
S(A)= \frac{\gamma_c}{\gamma} \frac{A}{4\ell_{\rm P}^2}-\frac{1}{2} \ln{\frac{A}{\ell_{\rm P}^2}+O(A^0)}\ ,
\ee
where $\gamma$ is the Barbero-Immirzi parameter \cite{BI} (a free real parameter in the theory)
and $\gamma_c$ a numerical constant obtained from the counting. Fixing $\gamma$ to be equal to
$\gamma_c$ ensures consistency with the Bekenstein-Hawking entropy
law for large areas. An important fact is that both definitions for the
horizon states to be considered agree with (\ref{SA}),
the only difference being the value of $\gamma_c$.

Alternatively, the complexity of the combinatorial
problem can be overcome by telling a computer to make an \emph{exact} counting by explicitly enumerating all states \cite{CQG}.
Though the exponentially growing number of states limits the counting to modest black hole sizes (a few hundred Planck areas), the results in this regime agree with the analytical computations in the large area limit.
Even more,
this direct computation reveals a richer behavior shown by
the spectrum when avoiding any approximation. The most degenerate quantum
configurations accumulate around certain evenly spaced values of
area, with a much lower degeneracy in the regions between those
values, thus giving rise to an effective
``quasidiscrete'' equidistant area spectrum, despite the fact that the area
spectrum in LQG is not equidistant. Furthermore,
this phenomenon is independent on the particular choice for the characterization
of the horizon degrees of freedom.
This result provides a contact point between LQG and the Bekenstein's
conjecture \cite{Bekenstein}
and has important
consequences for the physical properties of black holes, such as the
entropy, which displays an effective discretization \cite{PRL}, or
Hawking radiation, that could carry some quantum imprints
coming from the horizon structure at the Planck scale \cite{DF}.

This ``band structure'' arising in the black hole area spectrum of LQG calls for a more intuitive explanation, unraveling the origin of this phenomenon from the theory.
This is the main goal of the present paper. A recent work in this direction
has been done by Sahlmann in \cite{hanno}, where he gives some quantitative information about
the black hole area spectrum.
In this paper we will follow a rather different approach.
Despite the complexity of the combinatorial problem, which makes a
meticulous analysis unfeasible, the states can be properly handled by attending to a few
properties that allow us to obtain
the most relevant qualitative and quantitative information about the area spectrum, shedding some light on its behavior. In particular, this approach will help us to understand qualitatively the origin of the ``band structure'' and will also allow us to compute analytically the value of area corresponding to each ``peak'' of degeneracy.

We have organized the rest of the paper as follows. In sections (\ref{countings}) to (\ref{previous}) we review the previous works, paying special attention to the aspects related with our arguments, and establishing the notation we are going to use, while sections (\ref{ricor}) and (\ref{analysis}) contain the main body of the present work. Section (\ref{countings}) is devoted to set up the combinatorial problem. In section (\ref{sectionMDD}) some previous analytical results are presented. Section (\ref{previous}) contains a summary of the computational results that showed the behavior that we are going to analyze. We present our qualitative picture and our quantitative analytical computations in section (\ref{ricor}). The main results are analyzed in section (\ref{analysis}). We finally conclude with an outlook in section (\ref{conclusion}).

\section{Counting and labeling choices}
\label{countings}

In the Isolated Horizon (IH) framework in LQG black holes are treated
in an effective way, since they are introduced from the outset as an
inner boundary of the spacetime manifold before the quantization procedure
is carried out (see \cite{ABK, thiemann} for details). Isolated Horizon boundary conditions
are then imposed, which translate into quantum boundary conditions
after the quantization procedure. The horizon states are described by
a U(1) quantum Chern-Simons gauge theory, while
gravitational degrees of freedom of the bulk are represented by spin
networks, a set of edges with spin-like quantum numbers $(j,m)$ ($j\in\mathds Z/2, m=\{-j,-j+1,...,j\}$) that intersect to each other at
vertices. When an edge of the spin network pierces the horizon creating a puncture, it endows it with a ``quantum'' of area given by
\be
\label{areaspectrum}
a(j)= 8 \pi \gamma \ell_{\rm P}^2 \sqrt{ j(j+1)}\ ,
\ee
where $j$ is the corresponding label of the edge,
and with a quantum of curvature given by the label $m$ (since the Isolated
Horizon boundary conditions relate this label with the U(1) Chern-Simons
states on the horizon surface). Then, the quantum states of a
black hole with area $A$ must satisfy that
the sum of the contribution to the area from each puncture equals the
total horizon area,
\be
\label{areaconstrain}
A=8 \pi \gamma \ell_{\rm P}^2 \sum_{i=1}^p{\sqrt{j_i(j_i+1)}}\ ,
\ee
where $p$ is the number of punctures on the horizon.
Also a condition coming from the fact that the horizon is spherical must be imposed. This is
\be
\label{projconst}
\sum_i m_i=0\ ,
\ee
which is called
``projection constraint''. The problem of counting the black hole microstates
that account for its entropy is now reduced to a mathematically well defined
\emph{combinatorial} problem which can be stated as:

\emph{How many different configurations of labels distributed over a set of distinguishable\footnote{The fact that  punctures are distinguishable has its origin in some subtleties related with the action of diffeomorphisms during the quantization procedure, and plays a key role in the combinatorial problem.} punctures are there, for all possible finite numbers of punctures, such that the constraints} (\ref{areaconstrain}) \emph{and} (\ref{projconst}) \emph{are satisfied?}

There exists a certain ambiguity at this point, since there are
two proposals concerning which labels have to
be considered to account for all microscopic
configurations. The issue of which is the proper counting is, however,
beyond the scope of this paper, as the behavior that we want to analyze is obtained within
both of them.

The first of the two proposals was done by Domagala and Lewandowski
in \cite{DL} and was complemented by Meissner in \cite{M}. There, it is claimed that the horizon states are given by
punctures carrying only the $m_i$ labels (as these are the labels related
to the horizon states through the IH boundary conditions).
The constraint (\ref{areaconstrain}) is then reinterpreted in terms
of $|m_i|$.

The second proposal is due to Ghosh and Mitra \cite{GM}, and it
considers that both labels, $j_i$ and $m_i$, characterize the horizon quantum states.
In this case, both constraints (\ref{areaconstrain}) and (\ref{projconst}) can
be imposed as written above. The structure, results and main differences between both
models can be seen in Table \ref{counting comp}.

\begin{table}[t]

\caption{Comparison between the DLM and GM countings}\label{counting comp}
\begin{tabular}[t]{|c|c|c|}
\hline {}&DLM&GM\\
\hline Labels&$m_i$&$(j_i,m_i)$\\
\hline  Area&$8\pi\gamma \ell_{\rm P}^2\sum_i\sqrt{|m_i|(|m_i|+1)}$&$8\pi\gamma \ell_{\rm P}^2\sum_i\sqrt{j_i(j_i+1)}$\\
\hline Projection constraint&$\sum_i m_i=0$&$\sum_i m_i =0$\\
\hline Entropy&$S(A)=\frac{\gamma_{\rm DLM}}{\gamma}\frac{A}{4}-\frac{1}{2}\ln{A}$&$S(A)=\frac{\gamma_{\rm GM}}{\gamma}\frac{A}{4}-\frac{1}{2}\ln{A}$\\
\hline BI parameter&$\gamma=\gamma_{\rm DLM}=0.23753$&$\gamma=\gamma_{\rm GM}=0.27407$\\
\hline
\end{tabular}
\end{table}

For the purpose of this paper, we need to deal with the labels
related to area, so we will call this labels generically $s_i$, in such a
way that the $s_i$ will correspond to $|m_i|$ in the first case and
to $j_i$ in the second one. Furthermore, for the sake of simplicity,
we will deal only with integer numbers, so that we will take
$s_i=2|m_i|$ or $s_i=2j_i$ in each case. Then, in the DLM case,
there will be two possible values of $m_i$ for each $s_i$, namely
$\{-\frac{s_i}{2},\frac{s_i}{2}\}$, while in the GM one the
possible values of $m_i$ will be
$\{-\frac{s_i}{2},-\frac{s_i}{2}+1,...,\frac{s_i}{2}\}$, so there
will be $s_i+1$ values of $m_i$ for each $s_i$. This will be the
only difference that we will have to introduce in our arguments in
order to account for both counting models.

\section{Previous analytical results}
\label{sectionMDD}

In this section we review briefly the previous analytical
results on the counting of black hole
microstates \cite{ABK, DL, M, GM, GMdist} present in the literature.
When addressing the combinatorial problem described in the previous section,
a key point is to consider that punctures are distinguishable, as shown in \cite{ABK}.
With this in mind, one should consider all possible orderings of labels over punctures. But given a certain configuration of $s_i$ labels, all possible reorderings give rise to states with exactly the same area. One can then characterize a configuration just by fixing the number $n_s$ of punctures that take each particular value of $s$ and introducing all possible orderings as a certain degeneracy associated with this configuration. Thus,
in the remainder of the paper, a given set of numbers $\{n_s\}_{s=1}^{s_{max}}$ (where $s_{max}$ is the maximum value of $s$) will be called \emph{configuration}. A configuration will be permissible if it satisfies the constraint (\ref{areaconstrain}), which in terms of $n_s$ reads
\be
\label{area}
4\pi\gamma\ell_{\rm P}^2\sum_{s=1}^k{n_s\sqrt{s(s+2)}}=A\ .
\ee
Then, in order to consider all quantum states contained in a given configuration, one has to take into account the degeneracy coming from two sources:

\begin{itemize}
\item{one due to all possible reorderings of the $\{s_i\}$ labels over punctures,}
\item{and the other coming from all possible combinations of the $m_i$ labels associated to each configuration satisfying the constraint (\ref{projconst}).}
\end{itemize}

The difference between the two possible countings is contained in
this last term. For the only reason of being able to explicitly write down some
expressions, we are
going to consider for the moment the term corresponding to the GM
counting. One can then write the degeneracy associated
to a given configuration $\{n_s\}_{s=1}^{s_{max}}$ as:\footnote{The factor $(s+1)^{n_s}$
is the one accounting for all the possible values of $m_i$ associated to each
$s_i$, so in order to make the analysis for the counting of
\cite{DL} it will be enough to change this term by $2^{n_s}$}

\be
\label{degeneracy}
d(n_1,...,n_{s_{max}})=\frac{(\sum_s{n_s})!}{\prod_s{n_s!}}\prod_s{(s+1)^{n_s}}\ ,
\ee
where sums and products run from $s=1$ to $s_{max}$. In the above expression the
projection constraint is not being introduced, but this fact will not affect
the results that we are going to obtain in the remainder of the section.
This degeneracy was studied in
\cite{GM, GMdist}, where the question of which are the values
of $n_s$ that give rise to the maximal value of degeneracy, for a fixed value
of area, was addressed.

The degeneracy $d(n_1,...,n_{s_{max}})$ (or
equivalently $\ln d(n_1,...,n_{s_{max}})$) is maximized by varying $n_s$ subject to the
constraint (\ref{area}), which is introduced via a Lagrange
multiplier. This maximizing process can be easily worked out
in the \emph{large area limit}, where the variables $n_s\gg1$ can be
considered as continuous. The variational problem is
then easily solved by using Stirling's approximation, which gives the result\footnote{Although the equivalent expression obtained in \cite{M} was presented on the basis of different considerations, it can also be given the same interpretation of a degeneracy maximizing distribution.}
\be
\label{GM distribution prop}
\hat n_s=\frac{n_s}{\sum_s{n_s}}=(s+1)e^{-\lambda\sqrt{s(s+2)}}\ ,
\ee
where for consistency, $\lambda$ must satisfy the ``normalization condition''

\be
\label{normalization}
\sum_{s=1}^{s_{max}}{(s+1)e^{-\lambda\sqrt{s(s+2)}}}=1\ .
\ee

Numerical solutions of this equation, in the large area limit ($s_{max}\gg1$) gives $\lambda_{\rm GM}=0.861006$ ( or $\lambda_{\rm DLM}=0.746232$ in the case of the other counting proposal).

We will call this $\hat n_s$ distribution the
\emph{Maximal Degeneracy Distribution}\footnote{We will use the term \emph{distribution} as opposed to \emph{configuration}, in the sense that it gives the proportions between the different $n_s$ instead of the absolute value of each $n_s$. We will use this terminology in the next sections.} (MDD), and it will play a
pivotal role from now on. Besides, it was shown in \cite{GMdist,M} that
the introduction of the projection
constraint does not modify this distribution, so despite one starts without imposing it, the results can be considered as including this constraint.

\begin{figure}[htbp]
\begin{center}
\includegraphics[width=10cm]{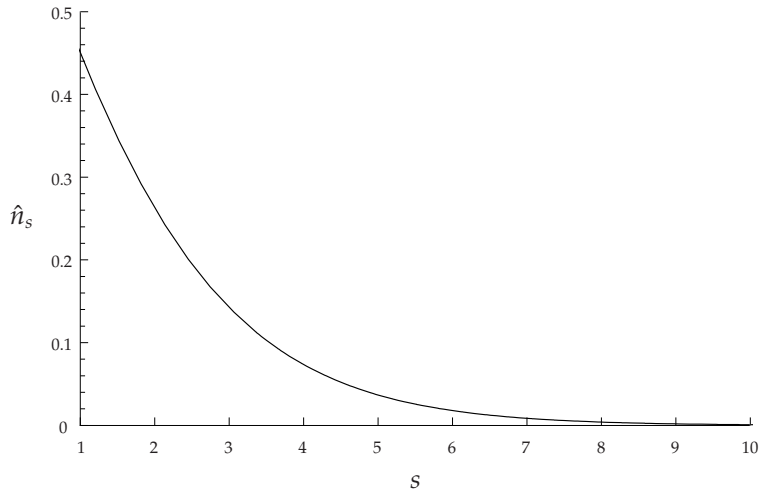}
\caption{The $\hat n_s$ given by the Maximal Degeneracy Distribution (MDD) is plotted as a function of $s$. The relevant contribution of the lower values of $s$ and the exponential decrease as $s$ grows are observed.}
\label{mdd}
\end{center}
\end{figure}

It is worth to note that, in the MDD, the proportions
between the different $n_s$ are maintained for different values
of area (the values of $n_s$ grow proportionally), and
then the values of $\hat n_s$ are independent of area. When plotting $\hat n_s$ (Figure \ref{mdd}), an interesting behavior is observed. Although the largest contribution comes
from the smallest value of $s$ (which contributes with approximately
one half of the punctures), the contribution of the next few values of $s$
is also significant. Nevertheless, the MDD shows an exponential decrease
as $s$ grows, so for $s$ larger than the smallest few values the contribution
will become negligible.

Once the MDD has been obtained, the total number of quantum states for a given value of
area can be computed. The result is \cite{M, GMdist}
\be 
d=\frac{\alpha}{\sqrt{A/\ell_{\rm P}^2}} e^{ \frac{\lambda}{4\pi\gamma\ell_{\rm P}^2}A} \ ,
\ee 
where $\alpha\sim O(1)$. It is seen that the number of quantum states grows
exponentially with area; the extra factor $A^{-1/2}$ appears due to the
introduction of the projection constraint (\ref{projconst}). From this
the entropy can be computed, obtaining

\be
\label{entropy result}
S(A)=\frac{\lambda}{\pi \gamma} \frac{A}{4\ell_{\rm P}^2}-\frac{1}{2} \ln{\frac{A}{\ell_{\rm P}^2}}+O(A^0)\ .
\ee

This result verifies the semiclassical Bekenstein-Hawking entropy formula
for large areas provided that $\gamma=\lambda/\pi$. Substituting the
value of $\lambda$ for each counting the corresponding values for
the Barbero-Immirzi parameter are obtained

$$\gamma_{\rm GM}=0.274066858 \ ,\ \ \gamma_{\rm DLM}=0.237532958\ .$$

\subsection{Large area limit}
\label{large area limit}

\begin{figure}[htbp]
\begin{center}
\includegraphics[width=10cm]{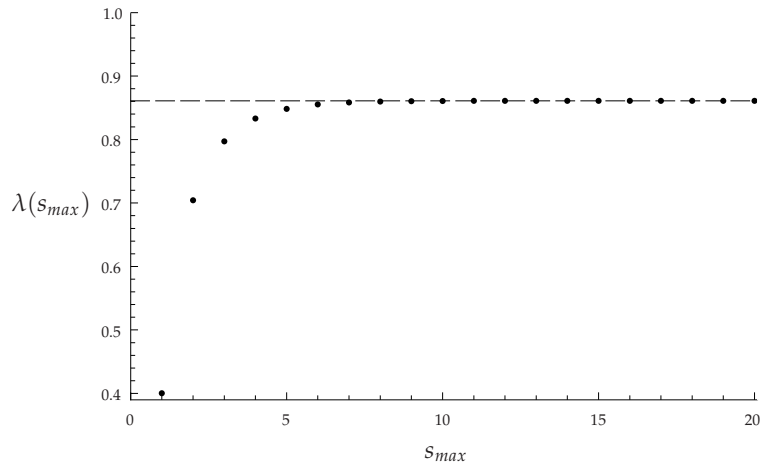}
\caption{The value of $\lambda$ as a function of $s_{max}$ is plotted, and compared with the asymptotic value.}
\label{lambda}
\end{center}
\end{figure}

In the previous computations the large area approximation was
involved. However, one can wonder about the meaning of ``large
area'' in this context. If one looks at the normalization condition
(\ref{normalization}), it is easy to see that the value of $\lambda$
obtained from it depends on the value of $s_{max}$
to which we are summing up. Then, as the value of $s_{max}$ depends
on the area, we have a $\lambda$ that is a function of area.
Nevertheless, if one studies the function $\lambda=\lambda(A)$ (or equivalently
$\lambda=\lambda(s_{max})$, as shown in Figure \ref{lambda}), one
sees that the value of $\lambda$ grows very quickly and saturates the
asymptotic value for relatively small values of $s_{max}$ (for $s_{max}$ around
$12$ the value of $\lambda$ only differs from the asymptotic value in a $0.006\%$). But this value of $s_{max}$ corresponds to values of area around $45 \ell_{\rm P}^2$. So for areas larger than that, we can say that we are already in the large area limit, as far as the
distribution (\ref{GM distribution prop}) is concerned.

\section{Previous computational results}
\label{previous}

In the previous section, some
approximations were employed in order to count the number of quantum
microstates compatible with a macroscopic black hole. One can legitimately be worried about
the fact that this approximations could be hiding part
of the richness of the problem. Fortunately, in spite of its
intrinsic complexity, an exact counting can be
performed to see whether there is a richer structure in the spectrum
or not. This can be done by means of an explicit enumeration
computational algorithm. The strategy is to generate systematically
all possible combinations of labels (for any possible number
of punctures), and to check one by one whether it satisfies the
required conditions. Then, by explicitly enumerating all states, one
can make an exact counting of the black hole quantum configurations
(for a given value of area) in this framework.
This was done in \cite{CQG}; here we are going to review
the main results obtained there and in subsequent work. Even when
such an exact counting can be done, the price to pay for overcoming
the complexity of the problem with an explicit enumeration is a
severe restriction to the black hole sizes that can be analyzed due to
the huge number of configurations to be counted. For that reason, the
available computing power allowed to analyze black holes up to just
a few hundred Planck area sizes.
However, these computations were enough to confirm the
results of the previous sections, namely the exponential growth of
the number of states with area and, when
imposing the projection constraint, the factor $A^{-1/2}$. The fact that this results are compatible
with the analytical computations gives one some confidence in the interest
of performing such a counting even though, due to computational limitations,
one is restricted to work in a small horizon area regime, far below the large area limit in which the Isolated Horizon framework in LQG was originally formulated.

\begin{figure}[htbp]
\begin{center}
\includegraphics[width=10cm]{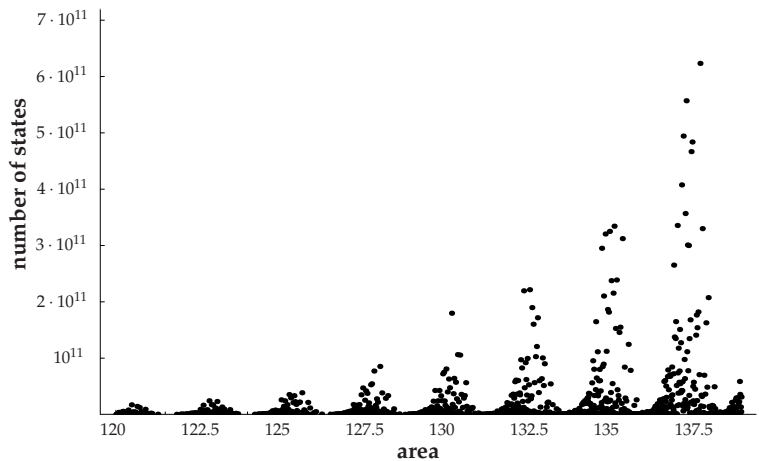}
\end{center}
\caption{Plot of the degeneracy (number of different horizon states in each area interval of $0.01\ell_{\rm P}^2$).
States accumulate around some equidistant values of area, exhibiting a band structure.}
\label{spectrum}
\end{figure}

But besides confirming the previous analytical results, the exact
counting showed a much richer behavior in the black hole area
spectrum \cite{PRL}. It was found that the black hole
quantum states are distributed according to a ``band
structure'' in terms of the area. The most degenerate configurations cluster around evenly spaced
values of area, giving rise to equidistant peaks of degeneracy,
with some orders of magnitude less degeneracy in the regions
between them (Figure \ref{spectrum}). This fact gives rise to an
\emph{effective} equidistant quantization of the black hole area in
LQG, even when the area spectrum in the theory
(\ref{areaspectrum}) is not equidistantly quantized. The most
relevant quantitative information about this phenomenon is the
fundamental area gap between peaks, which is given by
\be
\label{deltaA}
\Delta A= \gamma \chi \ell_{\rm P}^2 \ ,
\ee
where $\chi$ was estimated to be
$$\chi \approx 8.80\ .$$
A remarkable fact is that this result was obtained for both
choices of labels to be counted, and that all the difference resides just
in the value of the Barbero-Immirzi parameter.

\begin{figure}[htbp]
\begin{center}
\includegraphics[width=10cm]{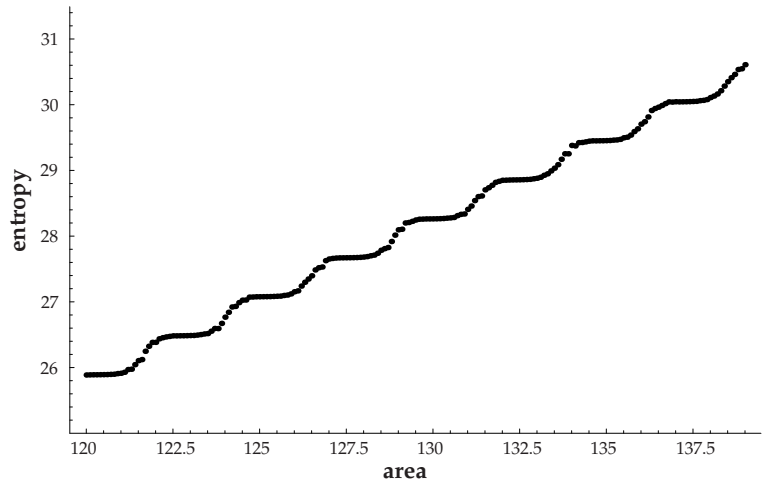}
\end{center}
\caption{Plot of the results for the entropy as a function of area (in Planck units) obtained with the computational counting.
The stair-like behavior, with a step width corresponding to $\Delta A$, is observed.}
\label{entropy}
\end{figure}

The obvious interest is now in the physical consequences of this structure.
The first clear consequence is in the entropy-area relation. This
periodic band structure in the area spectrum gives rise to a very
distinctive signal in the black hole entropy, namely a stair-like behavior of
entropy as a function of area,\footnote{For details on how to obtain the entropy shown in Fig.\ref{entropy} from the degeneracy of Fig.\ref{spectrum} see \cite{CQG,PRL}} as shown in Figure \ref{entropy}. Furthermore, the particular
structure of the area spectrum can also have some implications regarding the
black hole radiation spectrum, as pointed out in \cite{DF}.

On the other hand this regular pattern in the black hole area
spectrum provides a nice contact point with the heuristic ideas of
Bekenstein and Mukhanov \cite{Bekenstein} about
black hole area equidistant quantization. Even though the basic area
spectrum in LQG is not equidistant, this phenomenon shows that in
the case of black holes this equidistance in the spectrum appears in
a rather subtle way, namely as a result of the non trivial
degeneracy distribution. This point of contact becomes even more
intriguing when one realizes that the value of $\chi$ is close to $8 \ln 3\approx 8.788898$, as there is also a logarithmic constant arising from the heuristic considerations of Bekenstein and Mukhanov. It is evident that no reliable conclusion can be
extracted from this numerical proximity but it is worth keeping it in mind
to see if a more detailed work can confirm a deeper
relation behind this coincidence.


\section{The richness of discreteness}
\label{ricor}

In this section we seek to understand where the equidistant structure in the black hole spectrum comes from. We are going to analyze what happens to the MDD when one takes into account the discrete nature of the problem. Then, we are going to classify all configurations in sets characterized by two parameters, in such a way that the accumulation of states around the peaks of degeneracy becomes explicit and easy to study. Using these parameters, and some information extracted from the MDD, we will compute the value of area corresponding to each peak of degeneracy and then the area gap between peaks.

The first thing to consider is how is it possible to obtain information about the quasi-discrete structure of the spectrum using a distribution that was computed with approximations that seem to neglect all the information about this behavior. In this point the important thing to notice is that, in fact, the approximation that is hiding all the discrete information is to assume that one can find some configuration satisfying the MDD for any given value of area. When doing so, one is implicitly assuming that the $n_s$ numbers can take any possible value given by (\ref{GM distribution prop}), that in general are not integer values. It is clear that a non integer value for $n_s$ makes no sense. Then in order to find the actual maximally degenerate configuration, one should take the closest integer to each value of $n_s$ given by the MDD. However, if one modifies the value of $n_s$, one is modifying also its contribution to the area. Then, there are two possible cases, depending on the value of area we start with:

\begin{itemize}
\item{When one tries to find the closest integer configuration, the area changes of each $n_s$ compensate each other in such a way that at the end the integer configuration that we find takes almost the same value of area. Then we will be able to find some highly degenerate integer configurations with the same value of area we started from. This case would correspond with a peak of degeneracy.}
\item{When changing to the closest integer configuration, the deviations of each $n_s$ add up giving rise to a global area change so that the resulting integer configuration lies, in fact, in a different region of the spectrum. If one tries to find some integer configuration with the same value of area than the continuous one, it will not suffice to just take the closest integer to each $n_s$. One would be forced to modify considerably the $n_s$ distribution, in order to reach this value of area with integer $n_s$ values. But then, the obtained configuration will follow a distribution no longer close to the maximal degeneracy one, and would then have a much lower degeneracy. Therefore, one will not be able to find a highly degenerate integer configuration for this value of area. Such values of area are the ones corresponding to the regions of low degeneracy between peaks.}
\end{itemize}

Our task now is to find out which values of area correspond to the first case and which ones to the second. In order to do that, we will use a convenient classification of states.

\subsection{Classifying states}
\label{classify}

The combinatorial problem we are trying to address is a very complicated one, given the large number of variables (degrees of freedom) that come into play. For this reason, it is very difficult to handle all the information in a straightforward way. In order to be able to understand the underlying structure, we are going to organize all these configurations according to two parameters that will allow us to have a reasonable number of variables while keeping enough information for our analysis and computations. The two parameters we are going to consider to classify configurations are:

\begin{itemize}
\item{The number $p$ of punctures of the configuration, $p=\sum_{s=1}^{s_{max}}{n_s}$ , and}
\item{the sum over all punctures of the $s_i$ labels $S=\sum_{i=1}^p{s_i}=\sum_{s=1}^{s_{max}}{s\ n_s}$.}
\end{itemize}

For each pair of values of these parameters, we will have a set of many possible configurations. But the interesting thing is that if one fixes a given pair $(S,p)$, then the only freedom left to change the value of area associated to a configuration, is to distribute the $S$ ``units of $s$ label'' over the $p$ punctures in different ways (or in other words, to change the $n_s$ distribution). But the changes in area given by changing the distribution of $n_s$ are very small compared with the change in area given by modifying the parameters $S$ or $p$ in one unit (the requirement that all $n_s$ must be integer obviously implies that $S$ and $p$ can only take integer values). Then, by considering all possible $n_s$ distributions, one is covering an almost continuous region of area in the spectrum, while modifying $S$ or $p$ results in a discrete jump to another area region. Of course, if one modifies radically the $n_s$ distribution, from one extreme to the other, one can get changes in area larger than the one given by a change of one unit in $S$ or $p$, so these different area regions could overlap at some points.

On the other hand, although changing $S$ produces a jump in areas and so does a change in $p$, one could in principle modify both parameters in such a way that the final area change is small. In fact, as we are going to see, there is a way of changing $S$ and $p$ so that the area does not change. As pointed out in \cite{DF}, there is a very precise relation in the area spectrum of LQG that will help us to obtain this interesting relation between $S$ and $p$. One can check that the contribution to area given by one puncture with $s=6$ is exactly the same as the contribution given by four punctures with $s=1$. The interesting fact about this relation is that it is the only existing one for the low values of $s$ that are relevant to the highly degenerate configurations\footnote{The next exact relation is found between one puncture with $s=16$ and six punctures with $s=2$, but the contribution of punctures with $s=16$ to the highly degenerate configurations is completely negligible.} (as pointed out in section \ref{sectionMDD}, the value of $\hat n_s$ decreases exponentially with $s$ in the MDD).

Then, given a configuration, one can obtain another one with exactly the same value of area by removing a puncture with $s=6$ and adding four punctures with $s=1$ (decreasing the value of $n_6$ in 1 unit and increasing $n_1$ in 4 units). But this change implies increasing the number of punctures $p$ in three units and decreasing the sum of $s$ over all punctures ($S$) in two. Therefore, different pairs of parameters $(S,p)$ related by this transformation will be in the same area region.

We can write down this relation in a more concrete way. Given a value $S_0$ for $S$ and a value\footnote{Any value of $p$ larger than $3$ would be in correspondence with one of these tree values of $p_0$, i.e., a pair $(S,p=4)$ will correspond to $(S_0=S+2,p_0=1)$, and so on.} $p_0=1,2,3$ for $p$, all pairs $(S_t,p_t)$ that satisfy the following relation:
\be
\label{constant area Sp}
(S_t,p_t)=(S_0-2t,p_0+3t)\ ,
\ee
with $t\in\mathds Z$ such that $S_t\geq p_t$, are in the same region of area. $S_0$ will be the maximum value of $S$ and $p_0$ the minimum value of $p$ among all pairs $(S_t,p_t)$ satisfying this relation. Thus, if we consider the quantity $K=3S_0+2p_0$, we can associate to the same value of area all pairs of parameters satisfying

\be
\label{constant area}
3S+2p=K\ .
\ee

Then, for each value of $K$ we will obtain the configurations that appear in a certain region of area. In fact, it is important to notice that, if one takes into account the projection constraint, then the value of $S$ can only be even (for an odd value of $S$ would imply that $\sum_i{m_i}$ can only take half-integer values, and then it could not be zero). With this in mind, $K$ will only be allowed to take even values.

\subsection{Highly degenerate integer configurations}
\label{HDC}

Now, in order to account for the peaks of degeneracy, we need to consider the most degenerate integer configurations, which we will find with the help of the MDD. For a certain value of area, this distribution fixes a value for $S$ and $p$, that we will call $S_{md}(A)$ and $p_{md}(A)$. Furthermore, as the values of $\hat n_s$ are constant with area, $S_{md}(A)$ and $p_{md}(A)$ will grow proportionally with area, giving rise to a constant quotient $\hat s_{md}=\frac{S_{md}}{p_{md}}$. Hence, not all the pairs $(S,p)$ can contain maximally degenerate configurations; these configurations can only take values of $S$ and $p$ satisfying the quotient $\hat s_{md}$.
However, it is important to notice that the values of $S_{md}(A)$ and $p_{md}(A)$ fixed by the MDD are not, in general, integer numbers. Then, if starting from a configuration satisfying the MDD one changes to the closest integer values for each $n_s$ in order to find the actual maximally degenerate integer configuration, this would necessarily imply a change in $S$ and in $p$ to integer values. But as we have seen, to change $S$ and $p$ implies relatively large changes in area, unless such changes in $S$ and $p$ follow the ``constant area'' relation (\ref{constant area}). Therefore, if those $(S_{md}(A),p_{md}(A))$ satisfy this relation for an even value of $K$, it will be possible to find, for the same value of area, some integer pair $(S',p')$, with even $S'$, close to $(S_{md}(A),p_{md}(A))$, thus containing highly degenerate configurations. Otherwise, it will not be possible to find any highly degenerate configuration for that value of area, as explained at the begining of section \ref{ricor}.

In addition, among all configurations compatible with a given pair of values $(S',p')$, the most degenerate ones will be those having $n_s$ distributions close to the MDD, and therefore they will all appear together in a region of area much smaller than the total area covered by the set of all configurations with these values of $(S',p')$. Then, although the regions of area corresponding to different pairs $(S,p)$ can overlap (as pointed out above), the regions containing highly degenerate integer configurations will not. Thus, we expect to find the highly degenerate configurations clustered around some area values, each one corresponding to a different value of $K$.

We are going to compute in next section the values of area for which the MDD fixes a pair of values $(S_{md}(A),p_{md}(A))$ satisfying the constant area relation for each value of $K$, which will correspond to the values of area of the peaks of degeneracy.

\begin{figure}[htbp]
\begin{center}
\includegraphics[width=10cm]{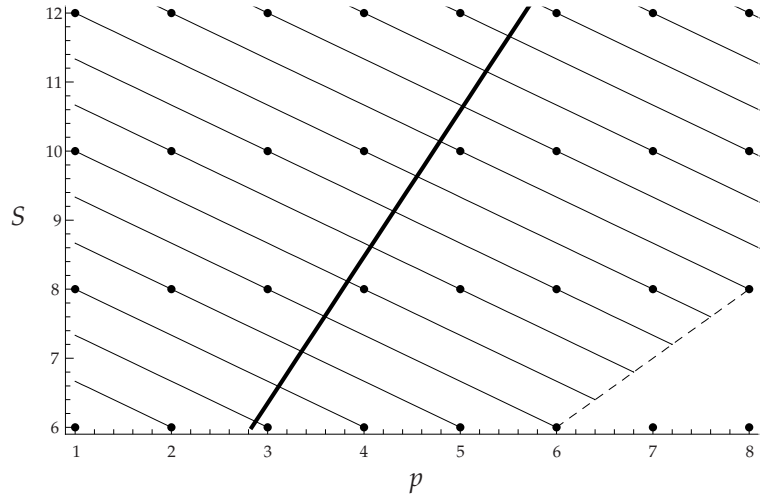}
\end{center}
\caption{Plot of the sum $S$ of spin labels vs. the number $p$ of punctures. All the discrete configurations are placed in the marked points. The thin lines represent ``constant area surfaces'' ($K$-lines) while the thick line cointains the values of $(S,p)$ that satisfy the quotient $\hat s_{md}$ (MD-line).}
\label{rectas}
\end{figure}

We can understand the above discussion in a more graphical way looking at Figure \ref{rectas}. The positive slope line represents the pairs $(S,p)$ that satisfy the maximal degeneracy quotient $\hat s_{md}$ (MD-line). Each of the negative slope lines represent the values $(S,p)$ related by (\ref{constant area}) for each even value of $K$ ($K$-lines). The marked points represent the allowed integer pairs $(S,p)$. Only for those points at where the MD-line intersects some of the $K$-lines, a close integer pair $(S',p')$ can be reached following the corresponding $K$-line (keeping area constant). In other words, the actual highest degeneracy configurations, that can only be found in the integer points close to the MD-line, correspond to the values of area associated to the point at which the corresponding $K$-line intersects this MD-line. The MDD provides the necesary information to compute the value of area associated to each point of the MD-line, as well as the slope of this MD-line. We can compute, therefore, the area of each intersection point and, thus, of the corresponding peak of degeneracy. Finally, computing the difference in area between two consecutive intersections (two consecutive even values of $K$) we will get the area gap between two peaks of degeneracy.



As a final remark in this section, we can analyze the effect of the projection constraint. In our model, this constraint is introduced by considering only even values of $K$ (i.e. even values of $S$). Then, if the projection constraint was not introduced, there would be an additional line between each two consecutive $K$-lines in Figure \ref{rectas}. This would correspond to having an additional peak of degeneracy between each two. But looking at Figure \ref{spectrum} one can see that, given the proportions of the spacing between peaks and the width of those peaks, placing an additional one between each two would almost result in no low degeneracy regions between them, hiding then the ``quasi-discrete'' structure of the spectrum. Then, as pointed out in \cite{hanno}, the regular pattern we are studying is a general feature that affects to all states and not only those satisfying the projection constraint. But it is precisely the introduction of this constraint what makes the ``discrete'' structure to arise in a clear and relevant way.

\subsection{Computation of $\Delta A$}
\label{computation}

Let us then proceed to the explicit computation of these quantities. The steps we are going to follow are:

\begin{itemize}
\item{In the first place, using the MDD we will compute the quotient $\frac{S_{md}}{p_{md}}$ for maximally degenerate states ($\hat s_{md}$).}
\item{From (\ref{constant area}) we will obtain an explicit relation $S=S(p,K)$.}
\item{We will use this explicit relation and the value of the quotient $\hat s_{md}$ to compute the number of punctures of the maximally degenerate state $p_{md}(K)$ for a given value of $K$ (the value of p at which the MD-line intersects a $K$-line in Figure \ref{rectas}).}
\item{Then, again using the MMD, we will compute the mean contribution to area $\hat A_{md}$ of a puncture in a configuration satisfying this distribution.}
\item{Thus, the value of area associated to an intersection with a line characterized by $K$ will be $A_{md}(K)=p_{md}(K)\hat A_{md}$.}
\item{Finally, computing the difference between $A_{md}(K)$ for two consecutive even values of $K$ ($A_{md}(K+2)-A_{md}(K)$), we will obtain the value of $\Delta A$.}
\end{itemize}

In order to compute $\hat s_{md}$ it is worth noticing that the quantity $\frac{S}{p}$ can be seen as the mean value of $s$ of each puncture in a configuration. Then, to compute this value in the case of the MMD we can write
\be
\label{MDslope}
\hat s_{md}=\sum_s{s\hat n_s}.
\ee
Thus we can compute the value of $\hat s_{md}$ and we know that
\be
\label{MDquotient}
\frac{S_{md}}{p_{md}}=\hat s_{md}\ .
\ee
Now, from the relation between $S$ and $p$ in a given band ($3S+2p=K$), we can extract the following equation
\be
\label{relationSp}
S(p,K)=\frac{K}{3}-\frac{2}{3}p\ .
\ee
Plugging this into (\ref{MDquotient}), we get
$$\frac{S_{md}(p,K)}{p_{md}(K)}=\frac{\frac{K}{3}-\frac{2}{3}p_{md}(K)}{p_{md}(K)}=\hat s_{md}\ ,$$
leading to
\be
\label{MDpunctures}
p_{md}(K)=\frac{K}{3\hat s_{md}+2}\ .
\ee
We have then the number of punctures that correspond to a maximal degeneracy configuration for a given value of $K$ (the intersection for a given $K$-line).

Now, to compute the mean contribution to the area from a puncture in a configuration satisfying the MDD, we proceed in the same way as we did to compute $\hat s_{md}$. We then write
\be
\label{MDarea}
\hat A_{md}=\sum_s{a(s)\hat n_s}=4\pi\gamma\ell_{\rm P}^2\sum_s{\hat n_s\sqrt{s(s+2)}}\ .
\ee
With this expression we can write the value of area associated to each of the intersections for each value of $K$,
\be
\label{peak area}
A_{md}(K)=p_{md}(K)\hat A_{md}=\frac{K\hat A_{md}}{3\hat s_{md}+2}\ .
\ee

We have then arrived at the main goal of the paper, i.e. obtaining the value of area associated to the corresponding peak of degeneracy for each value of $K$. With this expression we can compute numerically the value of area of each peak of degeneracy. We can also easily see that $A_{md}$ has a linear dependence on $K$, so the peaks are evenly spaced. We can hence compute this spacing just by taking the difference between two consecutive values of $K$,
\be
\label{delta}
\Delta A=A_{md}(K+2)-A_{md}(K)=\hat A_{md}(p_{md}(K+2)-p_{md}(K))=\frac{2\hat A_{md}}{3\hat s_{md}+2}\ .
\ee
Then, finally, writing explicitly all the terms in the above result, we can express the value of the area gap between peaks as

\be
\label{delta GM}
\Delta A_{\rm GM}=\chi_{\rm GM}\gamma_{\rm GM}=\frac{8\pi\gamma_{\rm GM}\ell_{\rm P}^2\sum_s{\sqrt{s(s+2)}(s+1)e^{-\lambda_{\rm GM}\sqrt{s(s+2)}}}}{3(\sum_s{s(s+1)e^{-\lambda_{\rm GM}\sqrt{s(s+2)}}})+2}\ .
\ee

At this point, we can recall that the only difference in all this discussion between the label choice we are using and the other comes from the degeneracy associated to the combinations of $m_i$ compatible with each configuration.
Introducing this change, the result for the $\Delta A_{\rm DLM}$ in the case of the counting of \cite{DL} is

\be
\label{delta DL}
\Delta A_{\rm DLM}=\chi_{\rm DLM}\gamma_{\rm DLM}=\frac{8\pi\gamma_{\rm DLM}\ell_{\rm P}^2\sum_s{2\sqrt{s(s+2)}e^{-\lambda_{\rm DLM}\sqrt{s(s+2)}}}}{3(\sum_s{2se^{-\lambda_{\rm DLM}\sqrt{s(s+2)}}})+2}\ ,
\ee

with the corresponding $\lambda_{\rm DLM}$. 


\section{Analysis of the results}
\label{analysis}

In this section we present the numerical values obtained for $\chi$ and analyze them. The resulting values of the expressions we have found can be easily computed using Mathematica\texttrademark{} and we get

$$\chi_{\rm GM}=8.789242\ ,\ \ \chi_{\rm DLM}=8.784286\ .$$
The fact that the difference between these two values is in the fourth digit gives us a hint on the level of accuracy that is being reached. Besides, it was pointed out in \cite{PRL} that the value of $\chi$ is numerically close to $8\ln{3}=8.788898$. One can see that the above results coincide, also up to the fourth digit, with this value, and furthermore, that the value of $8\ln{3}$ is contained between the two above values of $\chi$. One can compute the deviations between those three values:

$$\frac{|\chi_{\rm GM}-8\ln{3}|}{8\ln{3}}=0.000039=0.004\%\ ,$$
$$\frac{|\chi_{\rm DLM}-8\ln{3}|}{8\ln{3}}=0.00052=0.05\%\ ,$$
$$\frac{|\chi_{\rm GM}-\chi_{\rm DLM}|}{\chi_{\rm GM}}=0.00056=0.06\%\ .$$
Then, with a precision of $0.06\%$, the values of $\chi_{\rm GM}$, $\chi_{\rm DL}$ and $8\ln{3}$ are the same. Of course, this is still not a rigorous proof that $\chi$ is equal to $8\ln{3}$, but it is relevant to see how, when one improves the accuracy of the calculations, the numerical coincidence keeps being satisfied.

\begin{figure}[htbp]
\begin{center}
\includegraphics[width=10cm]{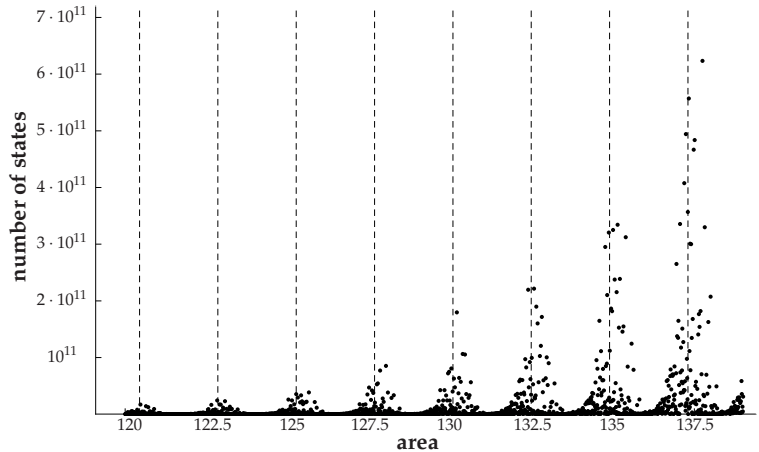}
\end{center}
\caption{Comparison between the analytical values for the area of the peaks (dashed vertical lines) and the actual peaks obtained from the computational data.}
\label{peak comparison}
\end{figure}

\begin{figure}[htbp]
\begin{center}
\includegraphics[width=10cm]{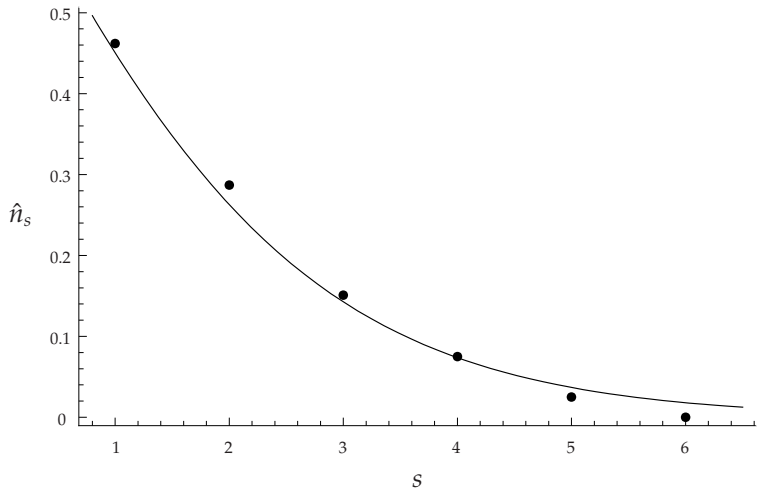}
\end{center}
\caption{The mean values of $\hat n_s$ obtained from the computational results (points) are compared with the exponential decrease expected from the MDD.}
\label{MDD comparison}
\end{figure}

Let us end this section with two remarks.

\begin{itemize}
\item{One can check whether the results obtained with the model
presented here are in good agreement with the computational data
obtained from the algorithm of \cite{CQG}. It can be seen in figure (\ref{peak
comparison}) how the values for the area of the peaks
that we obtained here fit the peaks observed in the spectrum
obtained from the computer. We see how the analytical values match
in a nice way with the computational data. On the other
hand, in figure (\ref{MDD comparison}) the mean values
of the $\hat n_s$ obtained with the computer for the five most degenerate configurations
of three consecutive peaks (with areas
between $170$ and $177\ell_{\rm P}^2$) are compared with those given by the MDD. One can check
that, even for this extreme low value of area, the agreement is
quite good, as expected from the analysis in section \ref{large area limit}, so one can feel confident with the use of the MDD in the computations. Finally, using these computational data we have observed that,
in fact, all configurations giving relevant contributions to the degeneracy at any given peak are characterized by pairs of values $(S,p)$ that satisfy the relation (\ref{constant area}) for the corresponding value of $K$, in complete agreement with the analysis presented in section \ref{ricor}.
Thus, the
computational data support the fact that the model presented here works reasonably
well.}
\item{At this point we can analyze the results previously obtained in \cite{hanno}. There, the problem is addressed using a rather different approach, namely, reformulating it in terms of the so called random walks. It is very interesting to see how, within this alternative approach, the accumulation of states around certain values of area also becomes manifest. By treating the area spectrum of LQG as an effectively quasi-equidistant spectrum, a way to compute the area gap between peaks is then proposed. The value for $\Delta A$ was obtained as $2/3$ of the spacing in this effectively quasi-equidistant area spectrum. This $2/3$ factor was introduced {\it ad hoc} in order to fit the computational data. A noteworthy fact is that this independent derivation gave rise to the same expresion (\ref{delta DL}).  Nevertheless, from the point of view of the authors it is not easy to reconcile the introduction of this $2/3$ coefficient with the qualitative picture of a quasi-equidistant spectrum being the origin of the observed regular pattern.

With the picture presented here, it is now rather easy indeed to
understand where this coefficient comes from.
The fundamental area gap in the quasi-equidistant spectrum of \cite{hanno}
corresponds to the mean
area change given by increasing $S$ in one unit in our formalism.
But as we have seen, $S$ only takes even values. Then the minimum
increment in $S$ must be two. Furthermore, if one comes back to the
relation between $S_0$, $p_0$ and $K$ ($K=2S_0+3p_0$) then one can
check that for each even value of $S_0$ there are three corresponding
values of $K$ (the ones corresponding to $p_0=1,2,3$) and then to
three peaks of degeneracy. Hence, the mean area change given by
increasing $S$ in two units corresponds to three times the area
gap between peaks. Thus, the fundamental gap of the quasi-equidistant spectrum in \cite{hanno} is nothing but three halves of the area gap between peaks $\Delta A$.}
\end{itemize}

\section{Conclusion and outlook}
\label{conclusion}

Let us summarize the results of the paper. We have analyzed the
combinatorial problem and we have qualitatively understood the
reason why the highest degenerate configurations can only appear
for some values of area and not for all of them. When the discrete nature
of the problem is taken into account, there are regions of area for which
the ``discrete configurations'' are not allowed to satisfy a distribution
close to the one that gives the maximal degeneracy in the continuous case, thus giving rise to the observed pattern in the black hole area spectrum.
We have also verified that the analysis is valid for both choices of labels, as the arguments presented here apply equally to both cases, so it seems now
rather natural that the analyzed behavior of
the spectrum appear with both counting procedures.
Finally, our analytical computations allowed us to obtain the values of area for which the peaks of degeneracy should appear and showed that these values are evenly spaced. In addition, the results match in a nice way with the computational data obtained in \cite{CQG}, thus indicating the validity of the model. From this, we have also been able to compute the analytical value of the corresponding parameter $\chi$ for both label choices and we have found that the results coincide up to a precision of $0.06\%$. Furthermore, we have checked out that, up to this improved precision, the surprising numerical coincidence with $8\ln{3}$ keeps holding.

There are still some important open questions. On the one hand, one may ask which are the sources of this $0.06\%$ deviation. Moreover, it would be very interesting to obtain an analytical proof for the conjectured value of $\chi=8\ln{3}$. On the other hand, although the area gap between peaks obtained with our model has no dependence on the area, one may be interested in knowing what would happen to the width of the bands for large areas. Whether this width increases with area, thus hiding the quasi-discrete behavior, or not, is also an interesting issue to be investigated.
A comprehensive analysis of the full combinatorial problem could shed light on some of these questions. But undoubtedly, the most important and interesting open question is to find a consistent physical interpretation to this intriguing behavior of quantum black holes.




\section*{Acknowledgements}

We would like to thank H.~Sahlmann for very interesting discussions about the content of the paper, and also T.~Pawlowski and G.~Mena-Marug\'an. We thank J.A.~de~Azc\'arraga, J.F.~Barbero, A.~Corichi, H.~Sahlmann and E.J.S.~Villase\~nor for a careful reading of the manuscript. We also thank J.~Navarro-Salas and J.~Olivert for their advice and encouragement. We finally thank specially A.~Corichi for his helpful guidance. 

IA thanks T. Jacobson for his hospitality at the University of Maryland. 
JD thanks the Institute for Gravitation and the Cosmos, Penn State University, for its hospitality and A.~Corichi also for his hospitality at the Instituto de Matem\'aticas, UNAM.

This work was in part supported by ESP2005-07714-C03-01, FIS2005-02761 and FIS2005-05736-C03-03 (MEC) projects and by CONACyT U47857-F grant. IA and JD thank MEC for support through the FPU (University Personnel Training) Program.

\end{document}